\documentclass[prb,reprint,twocolumn,showpacs,amsmath,amssymb]{revtex4-1}
\makeatletter

\usepackage[T1]{fontenc}
\usepackage{lmodern}
\usepackage{graphicx}
\usepackage{subfigure}
\usepackage{dcolumn}
\usepackage{bm}
\usepackage{amsmath}
\usepackage{sidecap}

\def\k{$\kappa_{\rm L}$}
\def\ke{$\kappa_{\rm e}$}
\def\t{t-PbTiO$_3$}

\begin{document}

\title{Estimates of the thermal conductivity and the thermoelectric properties of PbTiO$_3$ from first principles}
\author{Anindya Roy}
 \email{aroy21@jhu.edu}
\affiliation{
Materials Science \& Engineering, Johns Hopkins University,
Baltimore, MD 21218, USA
}

\date{\today}

\begin{abstract}
The lattice thermal conductivity (\k) of PbTiO$_3$ (PTO) is estimated using a combination of {\em ab initio} calculations and semiclassical Boltzmann transport equation. The computed \k\ is remarkably low, nearly comparable with the \k\ of good thermoelectric materials such as PbTe. In addition, a semiclassical analysis of the electronic transport quantities is presented, which suggests excellent thermoelectric properties, with a figure of merit $zT$ well over 1 for a wide range of temperature. For thermoelectric applications, the \k\ could be further reduced by utilizing different morphologies and compositions.
\end{abstract}

\maketitle
\marginparwidth 2.7in
\marginparsep 0.5in
\def\scr{\scriptsize}

PbTiO$_3$ (PTO) is a well studied perovskite ferroelectric, and is used extensively in the technologically important ferroelectric/piezoelectric ceramic PbZr$_x$Ti$_{(1-x)}$O$_3$ (PZT). PTO transitions from the cubic paraelectric (PE) phase to the tetragonal ferroelectric (FE) phase at temperature $T_c = $ 763~K. Recent computational work predicted PTO to be a promising $p$-type transparent conducting compound based on its electronic structure.\cite{Hautier13} Following doping to increase electrical conductivity, we can speculate about its thermoelectric behavior. The thermoelectric figure of merit at temperature $T$ is given by $zT = \sigma S^2 T/(\kappa_{\rm L} + \kappa_{\rm e})$, where $\sigma$ is the electrical conductivity, $S$ is the Seebeck coefficient, and \k\ and \ke\ are the lattice contribution and the electronic contribution to thermal conductivity. As a first approximation, the \ke\ of semiconductors is related to $\sigma$ as $\kappa_{\rm e} = L\sigma T$. The Lorenz number $L$ is a near-constant for many materials, which leaves little room for controlling \ke\ independent of $\sigma$. On the other hand, researchers have been pursuing many methods to reduce \k\ to increase $zT$.\cite{Snyder08,Gaultois14}

PTO has a relatively small unit cell which may not favor low \k. However, it has been observed that anharmonicity, such as that seen in thermoelectrics like PbTe, can decrease \k\ despite small lattice constants. PTO is used in superlattices and alloys -- morphologies that could enhance phonon scattering and thus reduce \k\ further. Despite this promise, only limited amount of experimental results are available on the topic of room-temperature and high-temperature behavior of \k\ in ferroelectrics including PTO.\cite{Mante1967, strukov_thermal_1994, tachibana_thermal_2008} Computational estimate of \k\ could illuminate some of these points. With this in mind, a semiclassical analysis of \k, as well as the electronic transport parameters of PTO are reported in this paper. The calculations are based on Boltzmann transport equations (BTE), which rely on {\em ab initio} results as the input. This study indicates a very low \k\ and high thermoelectric potential for PTO.  

Here DFT-based lattice dynamics methods are applied to generate the second-order interatomic force constants (IFC2), which are then used in the BTE solver (for phonons) to determine \k. The lattice dynamics methods assume harmonic forces on the atoms. If the forces have large anharmonic components, the resulting IFC2 will be less accurate, which in turn will make \k\ less precise. Thus, IFC2 from extreme anharmonic cases, or that from dynamically unstable structures cannot be used to determine the corresponding \k. For example, the cubic PTO structure for which DFT calculations predict imaginary phonon frequencies is outside the scope of the present study. However, tetragonal PTO (\t) shows less anharmonicity under the same calculation methods, and hence provides an alternate route to study \k\ from first principles.

To understand if the degree of anharmonicity in \t\ is within a reasonable limit, I checked available experiments on PTO and other materials. The example of PbTe, an incipient ferroelectric and a leading thermoelectric, is particularly relevant in this context. Strong anharmonic interaction between longitudinal acoustic (LA) and transverse optic (TO) modes has been determined as the {\em cause} of the exceptionally low \k\ of PbTe.\cite{an_ab_2008, delaire_giant_2011} DFT-based methods successfully estimated \k\ in PbTe despite such strong anharmonic interaction.\cite{tian_phonon_2012, skelton_thermal_2014} Anharmonicity in the phonon dispersion results of PTO has been studied experimentally.\cite{shirane_soft_1970, burns_lattice_1973, foster_anharmonicity_1993} By performing least-square analysis of the phonon frequencies near $T_c$, Freire and Katiyar concluded that anharmonicity in \t\ is small.\cite{freire_lattice_1988} Taken together, the observations on PbTe and PTO validate the use of the DFT-based methods to determine \k\ for \t. Additionally, in this work the phonon dispersion and \k\ are computed for different tetragonal structures of \t\ to study how structural changes affect \k. This approach is similar to the recent work on PbTe and other Pb-chalcogenides by Skelton and co-workers.\cite{skelton_thermal_2014} Different \t\ structures show modest variation in \k, according to the calculations presented here; and the range of \k\ closely follows the experimental \k. This conclusion for PTO, and similar conclusions for PbTe and the related systems~\cite{skelton_thermal_2014}  suggest that the DFT-based determination of \k\ could be a robust method, applicable to other ferroelectric/antiferroelectric materials.

The DFT calculations presented in this work were performed with {\scr Vienna Abinitio Simulation Package} ({\scr VASP}).\cite{kresse-vasp1, kresse-vasp2} Local density approximation (LDA)-based projector-augmented wave (PAW) pseudopotentials,\cite{Blochl-paw1994} which included the semicore $p$ electrons of Ti, were used for these calculations. The self-consistent calculations (SCF) and ionic relaxations with fixed lattice parameters had a plane-wave cutoff of 400 eV, whereas a higher cutoff of 520 eV was applied when optimizing cell parameters. A $8\times8\times8$ Monkhorst-Pack $k$-point mesh was used for the Brillouin zone integration. Forces on ions were converged to less than 0.001 eV/\AA. Scalar relativistic effects were included in these calculations, but the spin-orbit effect was left out. According to these calculations, relaxed cell parameter for cubic PTO is $a=3.881$ \AA. The optimized \t\ (referred to as the S$_2$ structure from now on) has $a=3.856$ \AA, and $c/a=1.046$. The cell parameters and the internal coordinates of atomic positions agree well with the previously reported values obtained from DFT-LDA calculations.\cite{garcia_1996,marton_phonon_2013}  Electronic density of states (eDOS) of the optimized cubic and tetragonal structures were determined via non self-consistent calculations with a $k$-point grid of $30\times30\times30$. The top valence bands of PTO (both in tetragonal and cubic structures) show strong dispersive character and are formed by the hybridization between Pb 6$s^2$- and O 2$p^6$-like orbitals, whereas the bottom of the conduction bands are relatively flat and have large contribution from Ti 3$d$-like orbitals. The eDOS calculated in this work agree with previous reports.\cite{piskunov_bulk_2004,Hautier13}

Besides S$_2$, the phonon dispersion and \k\ were determined for three other tetragonal structures, all with relaxed ionic positions. Two of these structures have $a=3.856$ \AA\ with $c/a=1.06$ (S$_1$) and 1.03 (S$_3$), whereas the third structure has $a=3.904$ \AA, and $c/a=$1.03 (S$_4$). The systematic shift in phonon frequencies was studied using these tetragonal structures, by varying the lattice parameter $a$ while the tetragonality ratio $c/a$ was kept fixed, and vice versa. Phonon dispersion plots were computed under stringent convergence criteria, using the {\scr Phonopy} code,\cite{phonopy} with {\scr VASP} as the DFT calculator. A set of finite displacement calculations on the $4\times4\times4$ supercells of \t\ (containing 320 atoms) produced the IFC2 used in this work. 
\begin{figure}
\includegraphics[width=3.0in]{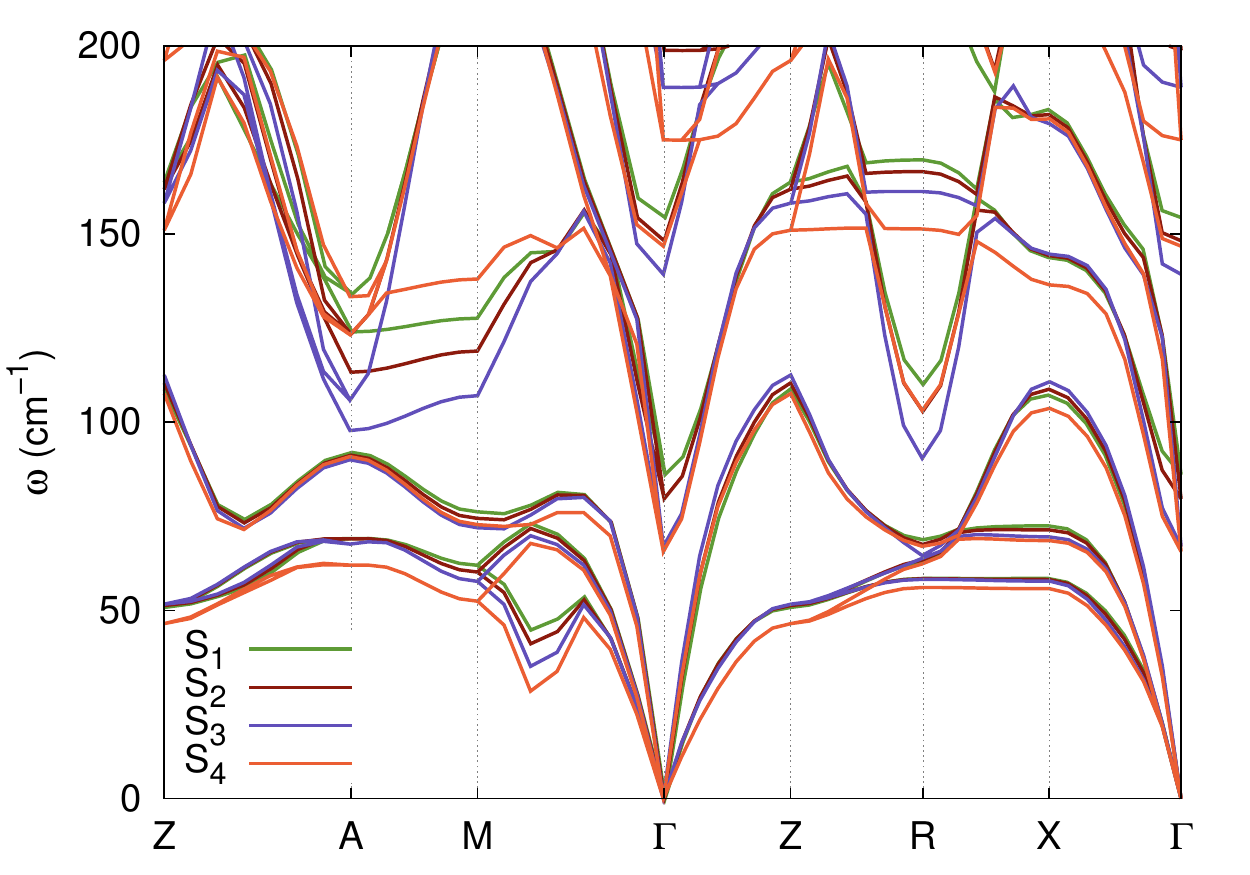}
\caption{\label{fig:phon-brnch} (Color online) Lowest branches in the phonon dispersion for the structures S$_1$--S$_4$.}
\end{figure}

Fig.~\ref{fig:phon-brnch} shows the phonon dispersion for the structures S$_1$--S$_4$. In this figure we zoom into the region with phonon frequencies up to 200 cm$^{-1}$ to inspect the thermal-current-carrying, low-lying phonon branches in greater detail. The acoustic modes of S$_4$ have smaller frequencies ("softer") than the rest, and those belonging to S$_1$--S$_3$ nearly overlap except in the $M$--$\Gamma$ region, where the structures with smaller $c/a$ have lower frequencies, i.e., $\omega$(S$_1$) $> \omega$(S$_2$) $> \omega$(S$_3$). Phonon frequencies of the lowest optic modes along the $A$--$M$--$\Gamma$ path follow the same pattern for S$_1$--S$_3$, whereas the optic modes related to S$_4$ appear to have {\em higher} frequencies than S$_1$--S$_3$. The lowest optic mode at $\Gamma$, known as $E$(TO$_1$),\footnote{conventional terminology, based on the decomposition of vibrational representation at $\Gamma$. See Ref.~\onlinecite{garcia_1996} for details.} stiffens for structures that are more tetragonal. This effect of tetragonality on the zone-center $E$(TO$_1$) has been observed in previous calculations.\cite{marton_phonon_2013} Experimental confirmation~\cite{burns_lattice_1973} of this trend of $E$(TO$_1$) is available via the temperature behavior of \t: a  rise in $T$ that reduces the tetragonality in \t\ is found to lower the associated $E$(TO$_1$) frequency. Thus the experiments and the calculations agree qualitatively.\footnote{In real life, strong anharmonic forces associated with the ferroelectric transition renormalize the phonon frequencies to real values, as PTO transitions to cubic structure.}

In general, a more quantitative connection between structure and $T$ can be determined in some cases using the quasiharmonic approximation (QHA), which includes the phonon contribution to Helmholtz free energy as a function of volume ($V)$ and $T$. Here we make a digression to look at the QHA analysis of \t\ before moving on to discuss \k\ calculations. QHA assumes harmonic forces on atoms at a specific volume, but allows for phonon frequencies to change with structure. Theoretical background and the implementation details of QHA are available in Refs.~\onlinecite{baroni_phonons_2001, skelton_thermal_2014}. Tetragonal symmetry in \t\ implies that the coefficient of thermal expansion (CTE), a second-rank tensor, has only two independent components, given by $\alpha_{xx} (= \alpha_{yy}$) and $\alpha_{zz}$ (following the Voigt notation). The components $\alpha_{xx}$ and $\alpha_{zz}$ are related to the principal components of strain as $\epsilon_{xx} = \alpha_{xx}\Delta T$, and $\epsilon_{zz} = \alpha_{zz}\Delta T$.\cite{nye1985} The volume CTE is $\alpha_{\rm V} = 2\alpha_{xx} + \alpha_{zz}$.  PTO shows negative thermal expansion (NTE) between $\sim$300~K--$T_c$.\cite{shirane_x-ray_1950,shirane-nte-1951,chen_thermal_2005} As $T$ rises towards $T_c$, the $c/a$ of \t\ decreases (as alluded to briefly in the last paragraph) while $a$ increases slightly, with an overall reduction in volume. Past attempts at applying QHA to determine the thermal expansion in PTO have run into difficulties. In one case, $\alpha_{\rm V}$ came out to be positive,\cite{wang_first-principles_2014} whereas another study reported a vastly overestimated magnitude of $\alpha_{\rm V}$.\cite{wang_first-principles_2013} In general, the lattice parameters of \t\ change in complex ways as functions of pressure,\cite{tinte_anomalous_2003, frantti_2007, ganesh_2009, zhu_pressure-induced_2014} which may be partly responsible for the difficulty in matching the QHA results for \t\ with the experimental CTE.

\begin{figure}
\includegraphics[width=3.0in]{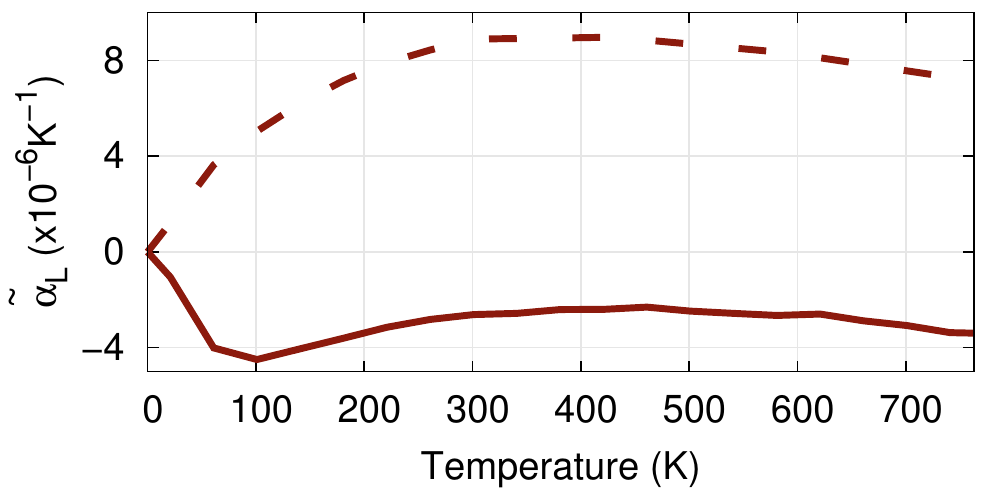}
\caption{\label{fig:t_exp} CTE components for \t\ as a function of temperature. The solid (lower) and the dashed (upper) curves represent $\tilde{\alpha}_{zz}$ and $\tilde{\alpha}_{xx}$, respectively.}
\end{figure}

In this work, the directly comparable components of CTE would be those determined under epitaxial constraints, because the phonon dispersion of different tetragonal structures (S$_1$--S$_4$) were compared by changing either the in-plane (x and y) or the out-of-plane (z) lattice direction(s), while holding the other fixed. The resulting CTE components can be termed {\em clamped} coefficients, $\tilde{\alpha}_{xx}$ and $\tilde{\alpha}_{zz}$, which are not the same as $\alpha_{xx}$ and $\alpha_{zz}$. The {\scr Phonopy-QHA} code\cite{phonopy} was used to compute $\tilde{\alpha}_{xx}$ and $\tilde{\alpha}_{zz}$. To determine $\tilde{\alpha}_{xx}$, a total of 12 structures were studied which spanned a variation in $a$ by $\pm 0.5\%$ around the optimized $a$ of \t\ (while the optimized $c$ was held fixed). Similarly, $\tilde{\alpha}_{zz}$ was obtained using 11 structures by varying $c$ in a range of $-0.3\%$ to $0.7\%$ around the optimized $c$, while $a$ remained fixed. In Fig.~\ref{fig:t_exp} we see that in the 300~K--600~K, $\tilde{\alpha}_{xx}$ is $\sim 9\times10^{-6}$~$^{\circ}$C$^{-1}$ and $\tilde{\alpha}_{zz}$ is $\sim -3\times10^{-6}$~$^{\circ}$C$^{-1}$. The phonon dispersion results for S$_1$--S$_4$ structures along with the computed values of $\tilde{\alpha}_{xx}$ and $\tilde{\alpha}_{zz}$ firmly establish that the structures with smaller $c/a$ ratio correspond to higher $T$ in these calculations.

Having explored how the S$_1$--S$_4$ structures relate to $T$, we now proceed to the results of the \k\ calculations. The {\scr ShengBTE} code\cite{ShengBTE_2014} used in this work iteratively solves BTE for phonons to determine \k. The IFC2 required in this code were obtained using the {\scr Phonopy} code as described earlier, generated with $4\times4\times4$ supercells via finite difference approach. The anharmonic IFC3 were obtained using the code {\scr thirdorder.py}\cite{thirdorder} (supplied with {\scr ShengBTE}), following the same finite difference methods on $3\times3\times3$ \t\ supercells. Interaction up to the third nearest neighbors was included in these calculations, and {\scr VASP} was used as the DFT engine. The linearized BTE was solved on a $\Gamma$-centered, $16\times16\times16$ $q$-point grid, which sufficiently converged \k. The supercells considered above to generate the set of IFC2 and IFC3 adequately converged \k. To test the convergence of \k\ on supercell size, additional sets of IFC2 and IFC3 were generated for the S$_2$ structure, on supercells of dimensions $3\times3\times3$ and $2\times2\times2$, respectively. Including the original choice ($4\times4\times4$ supercells for IFC2, and $3\times3\times3$ supercells for IFC3), four combinations of IFC2 and IFC3 were tested. For S$_2$, \k\ calculated with these four combinations lie within 0.2 Wm$^{-1}$K$^{-1}$, accurate enough for the purpose of this work.

\begin{figure}
\includegraphics[width=3.0in]{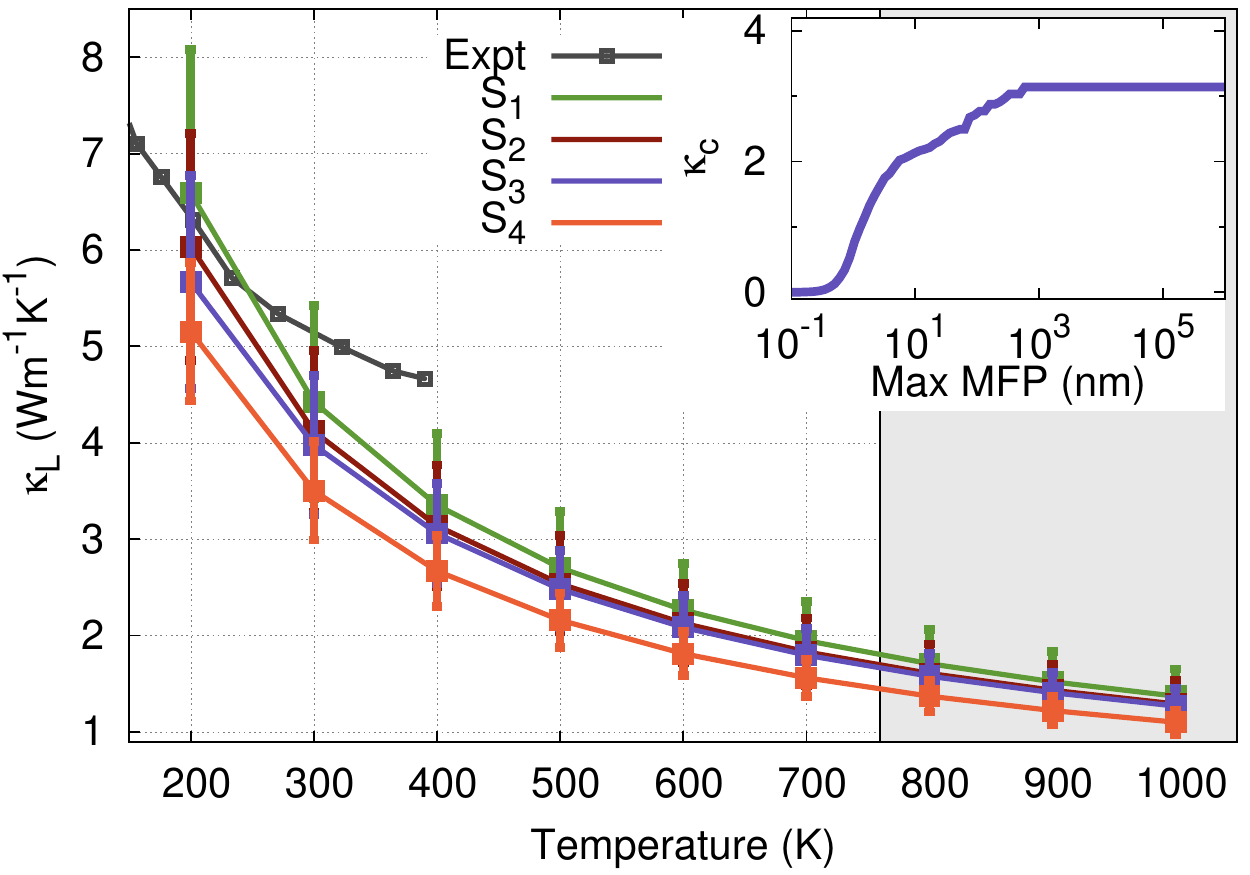}
\caption{\label{fig:kl-size} (Color online) \k\ as a function of $T$ for the S$_1$--S$_4$ structures of \t, along with the experimental data from Ref.~\onlinecite{tachibana_thermal_2008}. The upper bound and the lower bound of the bars represent the \k\ values along the in-plane and the out-in-plane axes, whereas the solid line represents one-third the trace of \k. The extrapolated region (for $T > T_c$) is shaded in grey. Cumulative \k\ is expressed as a function of maximum mean free path for the S$_2$ structure (inset).}
\end{figure}

Fig.~\ref{fig:kl-size} shows the \k\ values for the structures S$_1$--S$_4$, with the shaded region corresponding to extrapolated values of \k\ for $T > T_c$. First, we note that \k\ in PTO is low, especially at high temperature, and the structures with smaller $c/a$ show a lower \k. This effect is most pronounced for S$_4$, whereas the curves corresponding to S$_1$--S$_3$ show significant overlap. The structural dependence of \k\ in \t\ indicates that anharmonicity strongly affects \k. The experimental thermal conductivity results on perovskite ferroelectric samples~\cite{tachibana_thermal_2008, Mante1967} suggest that structural phase transitions do not significantly change the overall $T$ dependence of \k, except for a discontinuity close to $T_c$. This observation was used to calculate \k\ beyond $T_c$ using IFC2 and IFC3 corresponding to tetragonal structures. The top and the bottom end of the bars in Fig.~\ref{fig:kl-size} correspond to the \k\ values along the in-plane and the out-of-plane axes respectively, while the solid line represents the one-third the trace of the \k\ tensor. The inset of Fig.~\ref{fig:kl-size} shows the cumulative \k\ as a function of the maximum phonon mean free path for the optimized \t\ structure (S$_2$) at 400 K. According to these calculations, a nanoparticle of 10 nm diameter will have \k\ of about 2 Wm$^{-1}$K$^{-1}$ -- two thirds of its saturation value. Thus nanostructuring PTO may provide another route to reduce \k\ on top of its already low \k.

A comparison with the available experimental data for PTO shows overall agreement,\cite{tachibana_thermal_2008} although the computed \k\ appears underestimated in the 250~K--400~K. Experimental results include \ke, partly explaining the difference. Another contribution to this mismatch may be due to ignoring the spin-orbit effect in the calculations, reported to be important in certain cases.\cite{tian_phonon_2012} A third source of the difference could be a result of using harmonic approximation to calculate the phonon dispersion, which predicted (spuriously) smaller frequencies for the low-lying TO modes. This may have assigned a larger fraction of the thermal current to the TO modes in the present calculations. The optic modes in general have slower group velocity. Thus, a bigger fraction of the heat current carried by optic modes would lower \k\ overall.\cite{toberer_advances_2012} A conclusive insight would require observations from more experiments, and calculations that consider anharmonicity explicitly.

Low \k\ predicted for \t\ would make it a promising thermoelectric candidate if its electronic transport properties are good enough. To this end, semiclassical Boltzmann theory-based {\scr BoltzTraP}\cite{Madsen06} code was used to estimate $S$, $\sigma$, and \ke, within the constant scattering time approximation (CSTA). In CSTA, $\sigma$ and \ke\ are determined within a factor of the scattering time $\tau$, considered a parameter, whereas $S$ has no such dependence on $\tau$. Experimental results on $\sigma$, mobility and carrier concentration can be used to approximate $\tau$. However, such experimental results are lacking for conductive samples of PTO. As a substitute, a broad range of $\tau$, from 1~fs (=10$^{-15}$~s) to 25~fs, was used in these calculations.

\begin{figure}
\includegraphics[width=3.0in]{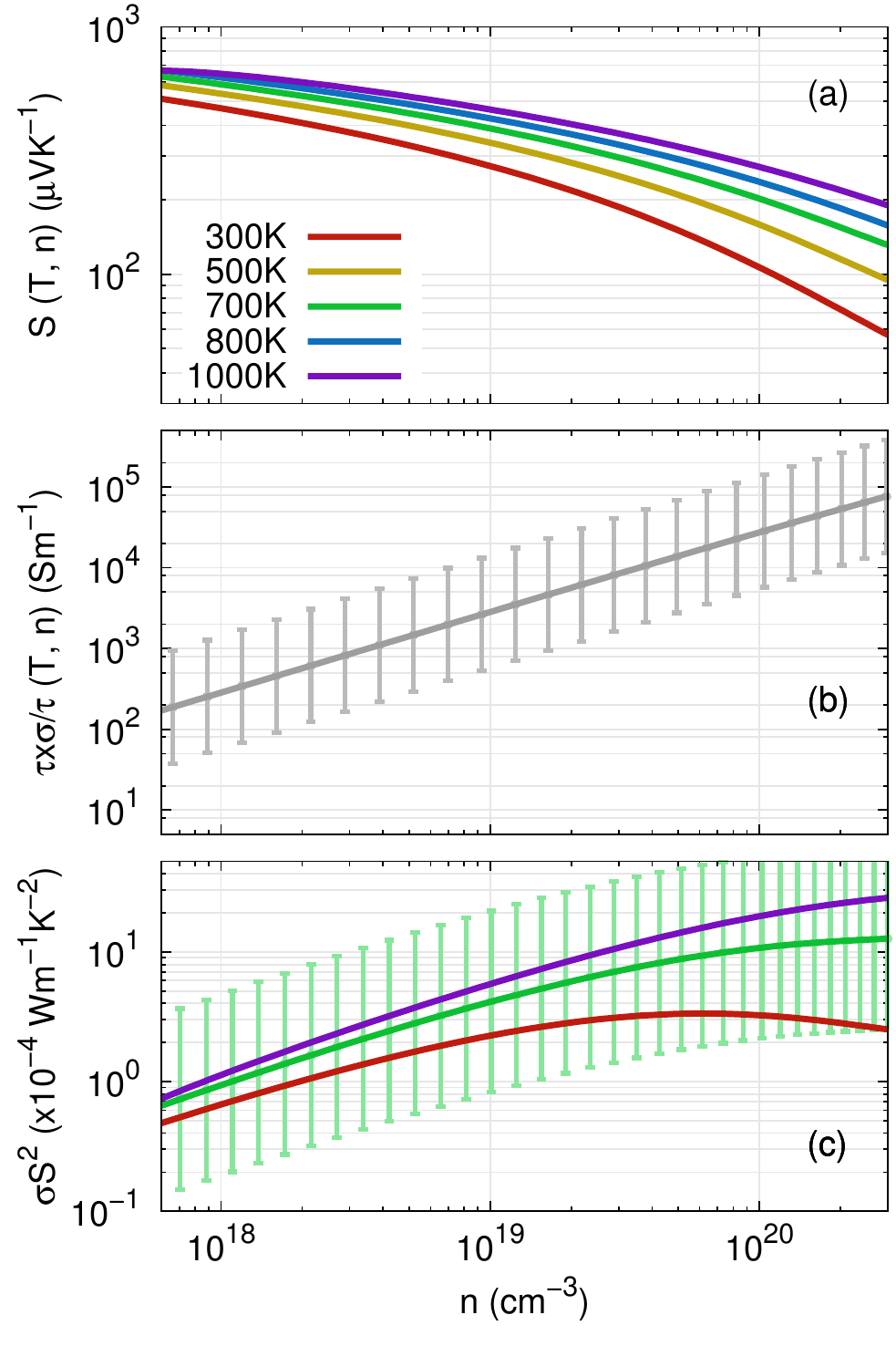}
\caption{\label{fig:therm-te} (Color online) Calculated thermoelectric parameter values: (a) $S(T,n)$ for five temperatures, (b) $\sigma/\tau\times\tau$ for three representative values of $\tau$: The solid line, and the upper and the lower limits of the vertical bars represent 5~fs, 25~fs, and 1~fs, respectively. (c) Power factor $\sigma S^2$ for systems at 300~K, 700~K, and 1000~K. The bars represent the same range of $\tau$ as in panel (b) for the 700~K curve [Same legends as (a)].}
\end{figure}
Fig.~\ref{fig:therm-te}(a)-(c) show the calculated values of $S$, $\sigma/\tau$, and the power factor $\sigma S^2$, all presented as functions of the carrier density $n$. In these calculations \ke\ was found to be small ($\le$0.5~Wm$^{-1}$K$^{-1}$ with $\tau =$ 5~fs) compared to \k, and hence ignored in further discussions. The DFT-derived eDOS were obtained from the cubic or the tetragonal structure, for $T \ge T_c$ and $T < T_c$, respectively; and were interpolated by a factor of 100 for the transport calculations. For the \t\ structure, one-third the trace of the $S$ and the $\sigma$ tensors are presented. As is widely observed for semiconductor thermoelectrics, $S$ increases with $T$ (for the same $n$), and drops with increasing $n$ (calculated at the same $T$). The $\sigma/\tau$ curve has no $T$ dependence, and in panel (b) we see a representative curve, with the bars spanning a region bound by $\tau =$ 25~fs as the upper limit, to $\tau =$ 1~fs as the lower limit. The solid line is drawn with $\tau =$ 5~fs. Fig.~\ref{fig:therm-te}(c) shows the estimated values of the power factor, which carries over the uncertainty in $\tau$ (the range of $\tau$ shown only for the $T=$ 700~K case, the rest drawn for $\tau = 5$~fs only). This plot indicates that $\sigma S^2$ is often greater than 0.002~Wm$^{-1}$K$^{-2}$ for a wide range of values of $\tau$ and $n$. For reference, a power factor of 0.002~Wm$^{-1}$K$^{-2}$ at 1000~K, accompanied by the estimated \k\ value of 1.15~Wm$^{-1}$K$^{-1}$, indicates a $zT$ of 1.7 -- highly promising as an initial estimate.

To summarize this work, the calculations presented here predict very low \k\ for PTO, influenced by the anharmonicity. The results largely agree with experiments. The low \k\ and the accompanying favorable electronic transport values indicate great promise of PTO as a thermoelectric material, provided the electrical conductivity is raised by doping. PTO and related perovskite functional materials have been synthesized and characterized in a variety of morphologies and compositions. Thus, processes such as alloying or increasing the complexity of the unit cell, among other measures, could be utilized to further reduce \k. Synthesizing PTO in disordered structure may provide another route to reduce \k\ while maintaining high $\sigma$. The top of the valence band of PTO is formed by the hybridization between the Pb $6s$ and the O $2p$ orbitals. The expansive, anisotropic Pb $6s$ orbitals may allow band-like hole conduction even in the absence of crystalline order, an idea originally suggested in the context of $n$-type amorphous transparent conducting oxides.\cite{Hosono06} This idea has recently been proposed as a means to reduce \k\ in ZnO-based $n$-type thermoelectrics.\cite{Roy-15-thermo} The present work demonstrates that \k\ can be computed for the low-symmetry, dynamically stable phases of ferroelectric/antiferroelectric materials, which can then be extended by reasonable approximation to high-$T$ regime. Comparison with experiments show that this approach can be a practical way to estimate \k\ on dynamically unstable systems which are otherwise intractable.

{\em The calculations related to this work will be available for download after the peer-review process.
}

\acknowledgments
The author is grateful to Michael L.~Falk for the discussions and the support, to Jes\'{u}s Carrete for the discussions regarding the implementation of {\scr ShengBTE}, and to Olivia Alley for the suggestions on the manuscript.
The work was supported by NSF grant DUE-1237992. This project used the computational resources of Extreme Science and Engineering Discovery Environment (XSEDE), which is supported by NSF grant number ACI-1053575; and The Maryland Advanced Research Computing Center (MARCC) supported by the State of Maryland.
\bibliography{pto-ttr}

\end{document}